\documentclass[12pt]{article}
\usepackage{graphicx}
\usepackage{amssymb}
\begin{document}
\title{%\vspace*{-3.5cm} 
Magnetic-field Manipulation of Chemical Bonding in Artificial 
Molecules
}
\author{%
Constantine Yannouleas\thanks{email: %
Constantine.Yannouleas@physics.gatech.edu}% 
~~and Uzi Landman\thanks{email: Uzi.Landman@physics.gatech.edu}\\
School of Physics, Georgia Institute of Technology,\\
Atlanta, GA 30332-0430
}
\date{August 2001}
\maketitle

\begin{abstract}

\noindent
The effect of orbital magnetism on the chemical bonding of lateral,
two-dimensional artificial molecules is studied in the case of a
2$e$ double quantum dot (artificial molecular hydrogen). It is found
that a perpendicular magnetic field reduces the coupling (tunneling) between 
the individual dots and, for sufficiently high values, it leads to
complete dissociation of the artificial molecule. The method used is building
on L\"{o}wdin's work on Projection Operators in Quantum Chemistry; it
is a spin-and-space unrestricted Hartree-Fock method in 
conjunction with the companion step of the restoration of spin and space 
symmetries via Projection Techniques (when such symmetries are broken).
This method is able to describe the full range of couplings in 
two-dimensional double quantum dots, from the strong-coupling regime 
exhibiting delocalized molecular orbitals to the weak-coupling and 
dissociation regimes associated with a Generalized Valence Bond combination 
of atomic-type orbitals localized on the individual dots. 
\end{abstract}
~~~~~~~~\\

\noindent
PACS: {73.21.La} {Quantum dots} --
      {85.35.-p} {Nanoelectronic devices} --
      {31.15.Rh} {Valence bond calculations} 

\newpage

\section{Introduction}
\label{intro}
Two-dimensional (2D) Quantum Dots (QD's) \cite{qds,cha} are usually referred 
to as artificial atoms, a term suggestive of strong similarities between 
these manmade nanodevices and the physical behavior of natural atoms. As a 
result, in the last few years, an intensive theoretical 
effort [2$-$13]
%\cite{cha,pfa,haw,vig,win,leb,man,mak,yl1,yl2,yl3,bar} 
has been devoted towards the elucidation of the appropriate analogies and/or 
differences. Recently, we showed [10$-$12]               %%%\cite{yl1,yl2,yl3}
that in the absence of a magnetic field the most promising analogies 
are mainly found outside the confines of the central-field approximation
underlying the Independent-Particle Model (IPM) and the ensuing physical
picture of electronic shells and the Aufbau Principle. Indeed, as a result of
the lower electronic densities in QD's, strong $e-e$ correlations can lead 
(as a function of the ratio $R_W$ between the interelectron repulsion and the
zero-point kinetic energy) to a drastically different physical regime, 
where the electrons become localized, arranging themselves in concentric 
geometric shells and forming electron molecules (referred to also as
Wigner molecules in analogy to Wigner crystallization \cite{wign} in infinite
media). In this context, it was found \cite{yl2,yl3} that the proper analogy 
for the particular case of a $2e$ QD is the collective-motion picture
reminiscent of the fleeting and rather exotic phenomena of the doubly-excited
natural helium atom, where the emergence of a ``floppy'' trimeric molecule
(consisting of the two localized electrons and the heavy $\alpha$-particle
nucleus) has been well established \cite{kel,ber}.

A natural extension [10,13,17$-$23]
%\cite{yl1,bar,nag,wen,burk,husa,yl4,pal,par} 
of this theoretical effort has also developed in the
direction of 2D QD Molecules (QDM's, often referred to as artificial
molecules), aiming at elucidating the analogies and differences between
such artificially fabricated \cite{wau,sch} molecular nanostrustures 
and the natural molecules. 
(Depending on the arrangement of the individual dots, two classes of
QDM's can be distinguished: lateral [10,13,17$-$21,24]
%\cite{yl1,bar,nag,wen,burk,husa,yl4,wau} 
and vertical \cite{pal,par,sch} ones.)

In a previous paper \cite{yl4} and for the case of zero 
magnetic field (field-free case),
we addressed the interplay of coupling and dissociation in lateral QDM's. We 
showed that this interplay relates directly to the nature of the coupling in 
the artificial molecules, and in particular to the question whether such 
coupling can be described by the Molecular Orbital (MO) Theory or the Valence 
Bond (VB) Theory in analogy with the chemical bond in natural molecules.

A major attraction of QD's and QDM's is the fact that, due to their larger 
size, orbital magnetic effects become important for magnetic-field values
easily attainable in the laboratory. This contrasts with the case of natural 
atoms and molecules for which magnetic fields of sufficient strength for the 
production of phenomena related to orbital magnetism are known to occur only 
in astrophysical environments, e.g., on the surfaces of neutron stars 
\cite{rud,lai}. In this paper, we study the effect of orbital magnetism on the
interdot coupling in a lateral, two-electron double quantum dot. This 
nanodevice represents an artificial analog to the natural hydrogen molecule 
and can be denoted as H$_2$-QDM. It has been suggested \cite{burk} that it can
function as the elemental two-qubit logic gate in quantum computing.
 
We find that the interplay of the MO versus the VB 
description provides the proper framework for understanding the 
influence of orbital magnetism on the chemical bonding of the H$_2$-QDM.
In particular, we show that a perpendicular magnetic field reduces 
the coupling between the individual dots and, for sufficiently high values, it
leads to the dissociation of the artificial molecule. As a result, in addition
to the obvious parameters of interdot barrier height and interdot separation, 
the magnetic field supplies a third variable able to induce dissociation
and thus to control the strength of the interdot coupling.

Our approach is twofold. As a first step, 
we utilize a self-consistent-field theory which can go beyond the
MO approximation, namely the spin-and-space unrestricted Hartree-Fock 
(sS-UHF), which was introduced by us \cite{yl1,yl2} for the 
description of the many-body problem of both single \cite{yl1,yl2} and 
molecular \cite{yl1} QD's. The equations used are given in Ref.\ \cite{solu},
where they are simply referred to as unrestricted Hartree-Fock (UHF); the
additional sS labeling employed by us emphasizes the range of possible 
symmetry unrestrictions in the solutions of these equations.
In particular, the sS-UHF differs from the more 
familiar restricted HF (RHF) in two ways: (i)
{\it it relaxes the double-occupancy requirement\/} $-$ namely, it employs 
different spatial orbitals for the two different (i.e., the up and down) spin
directions [thus the designation ``spin (s) unresricted''], and (ii) 
{\it it relaxes the requirement that the electron orbitals be constrained by
the symmetry of the external confining field\/} [thus the designation ``space
(S) unrestricted'']. Since it is a general property \cite{ring} of the HF 
equations to preserve at each iteration step the symmetries of the many-body 
hamiltonian (whenever they happen to be present in the HF electron density), 
the input trial density at the initial step must be constructed in such a way 
as to {\it a priori\/} reflect the relaxation of the two requirements 
mentioned above. Observe further that, in order to describe electron 
localization, the sS-UHF employs fully the fact that all $N$ (where $N$ is the
number of electrons) orbital-dependent effective (mean-field) HF potentials 
can be different from each other.

We remark 
that within the terminology adopted here, the simple designation Hartree-Fock 
(HF) in the literature most often refers to our restricted HF (RHF), in 
particular in atomic physics and the physics of the homogeneous electron gas. 
In nuclear physics, however, the simple designation HF most often refers to a 
space (S)-UHF. The simply designated unrestricted Hartree-Fock (UHF) as used 
in Chemistry (e.g., in calculations of open shell molecules) corresponds most 
often to our s-UHF (but not simultaneously space unrestricted HF).

As a second step, we will show that, in conjunction with L\"{o}wdin's
spin-projection technique \cite{low,paun},  the solutions with broken 
space symmetry allowed in QDM's by the sS-UHF provide a natural vehicle for 
formulating a Generalized Valence Bond (GVB) theory (see below) able to
describe the chemical bonding in artificial molecules in both the case of
an applied magnetic field and the field-free case.

\section{The two-center-oscillator confining potential}
\label{sec:2}

In the 2D two-center-oscillator\footnote{%
A 3D magnetic-field-free version of the TCO has been
used in the description of fission in metal clusters \cite{yl5}
and atomic nuclei \cite{grei}. } 
%**************   end footnote 1   *******************
(TCO), the single-particle levels 
associated with the confining potential of the artificial molecule are 
determined by the single-particle hamiltonian \cite{yl5,grei} %\cite{note1}
%\begin{equation}
\begin{eqnarray}
H=T &+& \frac{1}{2} m^* \omega^2_{x k} x^2
    + \frac{1}{2} m^* \omega^2_{y k} y^{\prime 2}_k \nonumber \\
    &+& V_{neck}(y) +h_k+ \frac{g^* \mu_B}{\hbar} {\bf B \cdot s}~,
\label{hsp}
\end{eqnarray}
%\end{equation}
where $y_k^\prime=y-y_k$ with $k=1$ for $y<0$ (left) and $k=2$ for $y>0$ 
(right), and the $h_k$'s control the relative well-depth, thus allowing studies
of hetero-QDM's. $x$ denotes the coordinate perpendicular to the interdot axis
($y$). $T=({\bf p}-e{\bf A}/c)^2/2m^*$, with ${\bf A}=0.5(-By,Bx,0)$, and the 
last term in Eq. (\ref{hsp}) is the Zeeman interaction with $g^*$ being the 
effective $g$ factor, $\mu_B$ the Bohr magneton, and ${\bf s}$ the spin
of an individual electron.
Here we limit ourselves to systems with $\hbar \omega_{x1}=\hbar \omega_{x2}=
\hbar \omega_x$. The most general shapes described by $H$ are two 
semiellipses connected by a smooth neck [$V_{neck}(y)$]. $y_1 < 0$ 
and $y_2 > 0$ are the centers of these semiellipses, $d=y_2-y_1$ is the 
interdot distance, and $m^*$ is the effective electron mass.

For the smooth neck, we use 
$V_{neck}(y) = \frac{1}{2} m^* \omega^2_{y k} 
[c_k y^{\prime 3}_k + d_k y^{\prime 4}_k] \theta(|y|-|y_k|)$, 
where $\theta(u)=0$ for $u>0$ and $\theta(u)=1$ for $u<0$.
The four constants $c_k$ and $d_k$ can be expressed via two parameters,
as follows: $(-1)^k c_k= (2-4\epsilon_k^b)/y_k$ and
$d_k=(1-3\epsilon_k^b)/y_k^2$, 
where the barrier-control parameters $\epsilon_k^b=(V_b-h_k)/V_{0k}$ 
are related to the actual (controlable) height 
of the bare barrier ($V_b$) between the two QD's, and 
$V_{0k}=m^* \omega_{y k}^2 y_k^2/2$ (for $h_1=h_2$, $V_{01}=V_{02}=V_0$).

The single-particle levels of $H,$
including an external perpendicular magnetic field $B$, 
are obtained by numerical diagonalization in a (variable-with-separation) 
basis consisting of the 
eigenstates of the auxiliary hamiltonian:
\begin{equation}
H_0=\frac{{\bf p}^2}{2m^*} + \frac{1}{2} m^* 
    \omega_x^2 x^2
      + \frac{1}{2} m^* \omega_{yk}^2 y_k^{\prime 2}+h_k~.
\label{h0}
\end{equation}
This eigenvalue problem is separable in $x$ and $y$, i.e., the
wave functions are written as $\Phi_{m \nu} (x,y)= X_m (x) Y_\nu (y)$.
The solutions for $X_m (x)$ are those of a one-dimensional
oscillator, and for $Y_\nu (y)$ they can be expressed through the parabolic
cylinder functions \cite{yl5,grei} $U[\alpha_k, (-1)^k \xi_k]$, where
$\xi_k = y^\prime_k \sqrt{2m^* \omega_{yk}/\hbar}$, 
$\alpha_k=(-E_y+h_k)/(\hbar \omega_{yk})$, 
and $E_y=(\nu+0.5)\hbar \omega_{y1} + h_1$ denotes
the $y$-eigenvalues. The matching conditions at $y=0$ for the left and
right domains yield the $y$-eigenvalues and the eigenfunctions
$Y_\nu (y)$ ($m$ is integer and $\nu$ is in general real). 

In this paper, we will limit ourselves to symmetric (homopolar) QDM's,
i.e., $\hbar \omega_{x}=\hbar \omega_{y1}=\hbar \omega_{y2}=
\hbar \omega_0$, with equal well-depths of the left and right dots, i.e.,
$h_1=h_2=0$. In all cases, we will use $\hbar \omega_0 =5$ meV and
$m^*=0.067 m_e$ (this effective-mass value corresponds to GaAs).

\section{The Many-Body Hamiltonian}
\label{sec:3}
The many-body hamiltonian ${\cal H}$ for a dimeric QDM comprising $N$ 
electrons can be expressed as a sum of the single-particle part $H(i)$ defined
in Eq.\ (\ref{hsp}) and the two-particle interelectron Coulomb repulsion,
\begin{equation}
{\cal H}=\sum_{i=1}^{N} H(i) +
\sum_{i=1}^{N} \sum_{j>i}^{N} \frac{e^2}{\kappa r_{ij}}~,
\label{mbh}
\end{equation}
where $\kappa$ is the dielectric constant and $r_{ij}$ denotes
the relative distance between the $i$ and $j$ electrons.

As we mentioned in the introduction, we will use the sS-UHF  
method for determining at a first level an approximate solution of the 
many-body problem specified by the hamiltonian (\ref{mbh}). 
The sS-UHF equations are solved in the Pople-Nesbet-Roothaan
formalism \cite{solu} using the interdot-distance adjustable basis formed with
the eigenfunctions $\Phi_{m \nu} (x,y)$ of the TCO defined in section 2.

As we will explicitly illustrate in section 4.2 for the case of the H$_2$-QDM,
the next step in improving the sS-UHF solution involves the use of 
Projection Techniques in relation to the UHF single Slater determinant.

\section{Artificial molecular hydrogen (H$_2$-QDM) in a magnetic field:
A Generalized Valence Bond approach}
\label{sec:4}

\subsection{The sS-UHF description}

As an introductory example to the process of symmetry breaking in HF,
we consider in this subsection the field-free ($B=0$) case of H$_2$-QDM
with $\kappa=20$ (this value is an intermediate one to the three different 
values of $\kappa$ that will be considered below in the case of an applied 
magnetic field). Fig.\ 1 displays the RHF and sS-UHF results for the 
$P=N_\alpha-N_\beta=0$ case (singlet) and for an interdot distance $d=30$ nm 
and an interdot barrier $V_b=4.95$ meV ($\alpha$ and $\beta$ denote up and 
down spins, respectively). In the RHF
(Fig.\ 1, left), both the spin-up and spin-down electrons occupy the same 
bonding ($\sigma_g$) molecular orbital. In contrast, the sS-UHF results 
exhibit breaking of the spatial reflection symmetry; namely, the
spin-up electron occupies an optimized\footnote{%
The optimized orbitals are anisotropic (i.e., non-circularly symmetric)
reflecting polarization effects due to the electronic interdot interaction.}
%*********  end footnote **************************
$1s$ atomic-like orbital
(AO) in the left QD, while the spin down electron occupies the 
corresponding $1s^\prime$ AO in the right QD. Concerning the total
energies, the RHF yields $E_{RHF}(P=0)=13.68$ meV, while the sS-UHF
energy is $E_{sSUHF}(P=0)=12.83$ representing a gain in energy of 0.85
meV. Since the energy of the triplet is $E_{sUHF}(P=2)=E_{sSUHF}(P=2)=
13.01$ meV, the sS-UHF singlet conforms to the requirement \cite{matt} that
for two electrons at zero magnetic field the singlet is always the
ground state; on the other hand the RHF MO solution fails in this respect. 

%***************** begin figure 1 **************************
\begin{figure}[t]
%\centering\includegraphics[width=7.8cm]{qds_lowdin_fig1j.eps}\\
%~~~~\\
\vspace{7cm}
\centering{\Large{\bf FIGURE 1}}
\vspace{7.5cm}
\caption{
Lateral H$_2$-QDM at zero magnetic field: Occupied orbitals (modulus 
square, bottom half) and total charge (CD) and spin (SD) 
densities (top half) for the $P=0$ spin unpolarized 
case. Left column: RHF results. Right column: sS-UHF results exhibiting
a breaking of the space symmetry. The numbers displayed with each orbital 
are their eigenenergies in meV, while the up and down arrows indicate an 
electron with an up or down spin. The numbers displayed with the charge 
densities are the total energies in meV. Unlike the RHF case, the spin density
of the sS-UHF exhibits a well developed spin density wave. 
Distances are in nm and the electron densities in $10^{-4}$ 
nm$^{-2}$. The choice of parameters is: $m^*=0.067 m_e$, $\hbar \omega_0=5$ 
meV, $d=30$ nm, $V_b=4.95$ meV, $\kappa=20$.
}
\end{figure}
%***************** end figure 1 **************************

\subsection{Projected wave function and restoration of the broken symmetry}

To make further progress, we utilize the spin projection technique to
restore the broken symmetry of the sS-UHF determinant (henceforth we will
drop the prefix sS when referring to the sS-UHF determinant),
\begin{eqnarray}
\sqrt{2}\Psi_{UHF}(1,2) &=& 
\left|
\begin{array}{cc}
u({\bf r}_1) \alpha(1) \; & \; v({\bf r}_1) \beta(1) \\
u({\bf r}_2) \alpha(2) \; & \; v({\bf r}_2) \beta(2) 
\end{array}
\right|  \nonumber \\
&\equiv& | u(1) \bar{v}(2) >~,
\label{det}
\end{eqnarray}
where $u({\bf r})$ and $v({\bf r})$ are the $1s$ (left) and $1s^\prime$ (right)
localized orbitals of the sS-UHF solution. An example of such orbitals for
the field-free case are displayed in the right column of Fig.\ 1. Similar
localized orbitals appear also in the $B \neq 0$ case, so that in general
the functions $u({\bf r})$ and $v({\bf r})$ are complex. 
$\alpha$ and $\beta$ denote the up and down spin functions, respectively. In 
Eq.\ (\ref{det}) we also define a compact notation for the $\Psi_{UHF}$ 
determinant, where a bar over a space orbital denotes a spin-down electron;
absence of a bar denotes a spin-up electron.

$\Psi_{UHF}(1,2)$ is an eigenstate of the projection $S_z$ of the
total spin ${\bf S} = {\bf s}_1 + {\bf s}_2$, but not of $S^2$.
One can generate a many-body wave function which is an
eigenstate of $S^2$ with eigenvalue $s(s+1)\hbar^2$ by applying the following 
projection operator introduced by L\"{o}wdin \cite{low,paun},
\begin{equation}
P_s \equiv \prod_{s^\prime \neq s}
\frac{S^2 - s^\prime(s^\prime + 1) \hbar^2}
{[s(s+1) - s^\prime(s^\prime + 1)] \hbar^2}~,
\label{prjp}
\end{equation}
where the index $s^\prime$ runs over the quantum numbers of $S^2$.

The result of $S^2$ on any UHF determinant can be calculated with the
help of the expression,
\begin{equation}
S^2 \Phi_{UHF} = 
\hbar^2 \left[ (N_\alpha - N_\beta)^2/4 + N/2 + \sum_{i<j} 
\varpi_{ij} \right]
\Phi_{UHF}~,
\label{s2}
\end{equation}
where the operator $\varpi_{ij}$ interchanges the spins of electrons
$i$ and $j$ provided that their spins are different; $N_\alpha$ and
$N_\beta$ denote the number of spin-up and spin-down electrons, respectively,
while $N$ denotes the total number of electrons. 

For the singlet magnetic state of two electrons ($N=2$), one has 
$N_\alpha=N_\beta=1$, $S_z=0$, and $S^2$ has only the two quantum numbers 
$s=0$ and $s=1$. As a result,
%\newpage
\begin{eqnarray}
2 \sqrt{2} P_{0} \Psi_{UHF}(1,2) & = & (1 - \varpi_{12}) \sqrt{2} 
\Psi_{UHF}(1,2) 
\nonumber \\
& = & | u(1) \bar{v}(2) > - \;| \bar{u}(1) v(2) >~.
\label{prj0}
\end{eqnarray}
In contrast to the single-determinantal wave functions of the RHF and sS-UHF 
methods, the projected many-body wave 
function (\ref{prj0}) is a linear superposition of two Slater determinants, 
and thus represents a corrective step beyond the mean-field approximation.

Expanding the determinants in Eq.\  (\ref{prj0}), one finds the equivalent
expression
\begin{equation}
2 P_{0} \Psi_{UHF}(1,2)  = 
(u({\bf r}_1)v({\bf r}_2) + u({\bf r}_2)v({\bf r}_1)) \chi(0,0)~,
\label{hl1}
\end{equation}
where the spin eigenfunction is given by
\begin{equation}
\chi(s=0;S_z=0)=(\alpha(1)\beta(2) - \alpha(2)\beta(1))/\sqrt{2}~.
\label{spin0}
\end{equation}
Eq.\ (\ref{hl1}) has the form of a Heitler-London (HL) \cite{hel} 
or valence bond\footnote{%
The early empirical electronic model of valence was primarily developed by 
G.N. Lewis who introduced a symbolism where an electron was represented by a 
dot, e.g., H:H, with a dot between the atomic symbols denoting a shared 
electron. Later in 1927 Heitler and London formulated the first quantum 
mechanical theory of the pair-electron bond for the case of the hydrogen 
molecule. The theory was subsequently developed by Pauling and others in the 
1930's into the modern theory of the chemical bond called the 
{\it Valence Bond Theory\/}. }
%**************   end footnote 2   *******************
\cite{coul,murr} wave function for the           %,murr,note2}
singlet magnetic state. However, unlike the HL scheme which uses the 
orbitals $\phi_L({\bf r})$ and $\phi_R({\bf r})$ of the separated 
(left and right) atoms,\footnote{%                             %\cite{note3}, 
Refs.\ \cite{burk,husa}
have studied, as a function of the magnetic field, the behavior of the 
singlet-triplet splitting of the H$_2$-QDM by diagonalizing the two-electron
hamiltonian inside the minimal four-dimensional basis formed by 
the products $\phi_L({\bf r}_1)\phi_L({\bf r}_2)$,
$\phi_L({\bf r}_1)\phi_R({\bf r}_2)$,
$\phi_R({\bf r}_1)\phi_L({\bf r}_2)$,
$\phi_R({\bf r}_1)\phi_R({\bf r}_2)$ of the $1s$ orbitals of the 
{\it separated\/} QD's. This Hubbard-type method \cite{burk} (as well as the
refinement employed by Ref.\ \cite{husa} of enlarging the minimal two-electron 
basis to include the $p$ orbitals of the separated QD's) is an improvement
over the simple HL method (see Ref.\ \cite{burk}), but apparently it is
only appropriate for the weak-coupling regime at sufficiently large distances 
and/or interdot barriers. In addition this method fails explicitly (it yields 
a triplet ground state at $B=0$ \cite{burk}) for small values of $\kappa$. Our
method is free of such limitations, since we employ here an interdot-distance 
adjustable basis (see section 2) of at least 70 spatial TCO molecular orbitals
when solving for the sS-UHF ones.
Even with consideration of the symmetries, this amounts to calculating
a large number of two-body Coulomb matrix elements, of the order of
$10^6$. }
%**************   end footnote 3   *******************
expression (\ref{hl1}) employs
the sS-UHF orbitals which are self-consistently optimized for any separation 
$d$, potential barrier height $V_b$, and magnetic field $B$.
As a result, expression (\ref{hl1}) 
can be characterized as a Generalized Valence Bond\footnote{%    %\cite{note4}
More precisely our GVB method belongs to a class of Projection Techniques
known as Variation {\it before\/} Projection, unlike the familiar in chemistry
GVB method of Goddard and coworkers [Goddard III, W.A.; Dunning, Jr., T.H.;
Hunt, W.J.; Hay, P.J. Acc. Chem. Res. 1973, 6, 368], which is a Variation 
{\it after\/} Projection (see Ref.\ \cite{ring}). }
%**************   end footnote 4   *******************
(GVB) wave function. Taking into account the normalization of the spatial 
part, we arrive at the following improved wave function for the singlet state 
exhibiting all the symmetries of the original many-body hamiltonian
(here, the spatial reflection symmetry is automatically restored along with
the spin symmetry),
\begin{equation}
\Psi^{\rm s}_{GVB}(1,2) = n_{\rm s} \sqrt{2} P_{0} \Psi_{UHF}(1,2)~,
\label{gvb}
\end{equation}
where the normalization constant is given by
\begin{equation}
n^2_{\rm s} = 1/(1 + S_{uv}S_{vu})~,
\end{equation}
$S_{uv}$ being the overlap integral of the $u({\bf r})$ and $v({\bf r})$ 
orbitals (which are complex in the presence of a magnetic field ${\bf B}$),
\begin{equation}
S_{uv}= \int u^*({\bf r})v({\bf r}) d{\bf r}~.
\end{equation}

The total energy of the GVB state is given by
\begin{equation}
E^{\rm s}_{GVB}=n^2_{\rm s}%
[h_{uu}+h_{vv} + S_{uv}h_{vu} + S_{vu}h_{uv}+J_{uv} + K_{uv}]~,
\label{engvb}
\end{equation}
where $h$ is the single-particle part (1) of the total hamiltonian (3),
and $J$ and $K$ are the direct and exchange matrix elements associated
with the $e-e$ repulsion $e^2/\kappa r_{12}$. For comparison, we give also
here the corresponding expression for the total energy of the HF ``singlet''
(i.e., the spin-contaminated determinant with $S_z=0$), either in the RHF 
($v=u$) or sS-UHF case, 
\begin{equation}
E^{\rm s}_{HF}=h_{uu}+h_{vv}+J_{uv}~.
\label{enghf}
\end{equation}

For the triplet with $S_z=\pm 1$, the projected wave function coincides with 
the original HF determinant, so that the corresponding energies in all
three approximation levels are equal, i.e., 
$E^{\rm t}_{GVB}=E^{\rm t}_{RHF}=E^{\rm t}_{UHF}$.

\subsection{Comparison of RHF, sS-UHF and GVB results}

In this section, we study in detail the behavior of the interdot coupling 
in the H$_2$-QDM in the presence of a perpendicular magnetic field.
We present numerical results for three values of the dielectric
constant, namely, $\kappa=45$ ($e-e$ repulsion much weaker than the case
of GaAs), $\kappa=25$ ($e-e$ repulsion weaker than the GaAs case), and 
$\kappa=12.9$ (case of GaAs). In particular we study the evolution
of the energy difference, $\Delta \varepsilon = E^{\rm s} - E^{\rm t}$, between
the singlet and the triplet states as a function of an increasing 
magnetic field varying from $B=0$ to $B=9$ T. The evolution of the two 
occupied sS-UHF orbitals of the singlet state is illustrated by plotting them 
at the two end values, $B=0$ and $B=9$ T. In all three figures (i.e., Fig.\ 2,
Fig.\ 3, and Fig.\ 4 below), the lower 
half corresponds to a vanishing interdot barrier $V_b=0$ (unified deformed 
dot at $B=0$), while the upper half corresponds to a finite value of $V_b$, 
being thus closer to the notion of a molecule proper. The interdot distance is
chosen to be $d=30$ nm in all three cases.

%***************** begin figure 2 **************************
\begin{figure}[t]
%\centering\includegraphics[width=12.5cm]{qds_lowdin_fig2j.eps}\\
%~~~~\\
\vspace{6.2cm}
\centering{\Large{\bf FIGURE 2}}
\vspace{6.5cm}
\caption{
Lateral H$_2$-QDM in the presence of a magnetic field and for $\kappa=45$: 
The left column displays the energy difference $\Delta \varepsilon$ between 
the singlet and triplet states according to the RHF (MO Theory, upper solid 
line), the sS-UHF (dashed line), and the GVB approach (Projection Method, 
lower solid line) as a function of the applied magnetic field $B$. The two
other columns display the sS-UHF spin-up ($\uparrow$) and spin-down 
($\downarrow$) occupied orbitals (modulus square) of the singlet state for 
$B=0$ (field-free case, middle column) and $B=9$ T (right column). The top
half corresponds to a bare interdot barrier of $V_b=3.71$ meV, while
the bottom half describes the no barrier case $V_b=0$.  
For $B=9$ T complete dissociation has been practically reached.
The choice of the remaining parameters is: $m^*=0.067 m_e$, $\hbar \omega_0=5$
meV, $d=30$ nm, and $g^*=0$. Distances are in nm and the orbital densities 
in $10^{-4}$ nm$^{-2}$. Insets: The overlap integral (modulus square) of the 
two orbitals of the singlet state as a function of $B$.
}
\end{figure}
%***************** end figure 2 **************************

\subsubsection{Case of $\kappa=45$}

Fig.\ 2 displays the evolution of $\Delta \varepsilon$ as a function of the
magnetic field for $\kappa=45$ and for all three approximation levels, i.e., 
the RHF (MO Theory, top solid line), the sS-UHF (dashed line), and the GVB 
(lower solid line). (The same convention for the $\Delta \varepsilon (B)$
curves is followed throughout this manuscript.) The case of $V_b=0$ is shown 
at the bottom panel, while the case of $V_b=3.71$ meV is displayed in the 
top panel. The insets display as a function of $B$ the overlap (modulus 
square, $|S_{uv}|^2$) of the two orbitals $u({\bf r})$ and $v({\bf r})$
of the singlet state. In this calculation, the effective $g$ factor was set 
equal to zero, $g^*=0$, so that the gain of energy due to the Zeeman 
effect does not obstruct for large $B$ the convergence of $\Delta \varepsilon$
towards zero. Due to its smallness relative to $\hbar \Omega_{eff}=
\hbar (\omega_0^2 +\omega_c^2/4)^{1/2}$ (where $\omega_c=eB/m^*c$ is
the cyclotron frequency), the actual Zeeman contribution can simply be added
to the result calculated for $g^*=0$.

We observe first that as a function of $B$, for both the $V_b=0$ and the 
$V_b=3.71$ meV cases and for all three levels of approximation, the 
$\Delta \varepsilon$ energy difference starts from a minimum negative value 
(singlet ground state) and progressively increases to zero; after crossing
the zero value, it remains positive (triplet ground state). However, for large
values of $B$ there is a sharp contrast in the behavior of the RHF curves 
compared to the sS-UHF and GVB ones. Indeed after crossing the zero axis, the 
RHF curves incorrectly continue to rise sharply and very early they
move outside the range of values plotted here (at $B=9$ T, the RHF values 
are 0.93 and 1.21 meV for $V_b=0$ and 3.71 meV, respectively). In
contrast, after reaching a broad maximum, the positive $\Delta \varepsilon$ 
branches of both the sS-UHF and GVB curves converge to zero for sufficiently 
large values of $B$. 

The convergence of the singlet and triplet total energies to the same
value indicates that the H$_2$-QDM dissociates as $B$ attains sufficiently
large values. This is also reflected in the
behavior of the overlaps (see insets) as a function of the magnetic field. 
In fact, the overlaps decrease practically to zero as a function of $B$,
suggesting that the two corresponding orbitals $u({\bf r})$ and $v({\bf r})$
of the singelt state tend to become strongly localized on the individual 
dots. 

The molecular dissociation induced by the magnetic field and the associated 
electron localization is further demonstrated by an examination of the orbital
densities (modulus square) themselves. These densities are plotted for
both the end values of $B=0$ (field-free case, middle column) and $B=9$ T
(strong magnetic field, right column). For $B=9$ T, it is apparent that the
plots portray orbitals well localized on the individual dots.
In contrast, for $B=0$ the orbitals are clearly delocalized over the
entire QDM. In particular, for $B=0$ and $V_b=0$ the two orbitals $u({\bf r})$
and $v({\bf r})$, which are different in general, have collapsed to the same 
1$s$-type distorted orbital associated with the single-particle picture
of a unified deformed dot. 

One effect of choosing a very weak $e-e$ repulsion ($\kappa=45$) is that in 
the $V_b=0$ case the three levels of approximation collapse to the same MO 
value (no symmetry breaking) for magnetic fields below $B < 2.9$ T, a fact 
that is also reflected in the orbitals themselves (see the $B=0$ case in the 
middle column of the lower half of Fig.\ 2). However, for $V_b=3.71$ meV, even
such a weak $e-e$ repulsion does not suffice to inhibit symmetry breaking in 
the field-free case. As a result, for $B=0$ and $V_b=3.71$ meV, the two 
$u({\bf r})$ and $v({\bf r})$ orbitals, although spread out over the entire 
molecule, are clearly different and the GVB singlet lies lower in energy 
than the corresponding MO value.

A second observation is that both the sS-UHF and the GVB solutions
describe the dissociation limit ($\Delta \varepsilon \rightarrow 0$) for
sufficiently large $B$ rather well. In addition in both the sS-UHF
and the GVB methods the singlet state at $B=0$ remains the ground state for all
values of the interdot barrier. Between the two singlets, the GVB one
is always the lowest, and as a result the GVB method presents an
improvement over the sS-UHF method both at the level of symmetry
preservation and the level of energetics. Furthermore, whether a singlet or
triplet, the GVB always results in a stabilization of the ground state;
the improved behavior of the GVB over the sS-UHF holds for all values of 
$\kappa$. 

We note that the failure of the MO (RHF) approximation to describe the
dissociation process induced by the magnetic field is similar to
its failure to describe dissociation of the molecule in the field-free
case as a function of the interdot barrier and distance (see Ref.\
\cite{yl4}). As discussed in Ref.\ \cite{yl4}, due to the underlying
MO picture, the spin-density functional calculations of Refs.\ \cite{nag,wen} 
also fail to describe the molecular dissociation process in the field-free
case.\footnote{% 
Symmetry breaking in coupled QD's within the LSD has been explored by
Kolehmainen, J.; Reimann, S.M.; Koskinen, M.; Manninen, M.
Eur. Phys. J. D 2000, 13, 731. However,
unlike the HF case for which a fully developed theory for the restoration
of symmetries has long been established (see, e.g., Ref.\ \cite{ring}),
the breaking of space symmetry within the 
spin-dependent density functional theory poses a serious dilemma 
[Perdew, J.P.; Savin, A.; Burke, K. Phys. Rev. A 1995, 51, 4531].
This dilemma has not been fully resolved todate; several remedies 
(like Projection, ensembles, etc.) are being proposed, but none of them 
appears to be completely devoid of inconsistencies [Savin, A. In 
Recent Developments and Applications of Modern Density Functional 
Theory, edited by Seminario, J.M.; Elsevier: Amsterdam, 1996, p. 327].
In addition, due to the unphysical self-interaction error, the 
density-functional theory is more resistant against symmetry breaking
[see Bauernschmitt, R.; Ahlrichs, R. J. Chem. Phys. 1996, 104, 9047]
than the sS-UHF, and thus it fails to describe a whole 
class of broken symmetries involving electron localization, e.g., the formation
at $B=0$ of Wigner molecules in QD's (see footnote 7 in Ref.\ \cite{yl1}),
the hole trapping at Al impurities in silica [Laegsgaard, J.; Stokbro, K.
Phys. Rev. Lett. 2001, 86, 2834], or the interaction driven
localization-delocalization transition in $d$- and $f$- electron systems,
like Plutonium [Savrasov, S.Y.; Kotliar, G.; Abrahams, E. 
Nature 2001, 410, 793]. }
%***************** end footnote 5 **************************
Such spin-density functional calculations are expected to fail
in the presence of a magnetic field as well.

%***************** begin figure 3 **************************
\begin{figure}[t]
%\centering\includegraphics[width=12.3cm]{qds_lowdin_fig3j.eps}\\
%~~~~\\
\vspace{6.2cm}
\centering{\Large{\bf FIGURE 3}}
\vspace{6.5cm}
\caption{
Lateral H$_2$-QDM in the presence of a magnetic field and for $\kappa=25$: 
The left column displays the energy difference $\Delta \varepsilon$ between 
the singlet and triplet states according to the RHF (MO Theory, upper solid 
line), the sS-UHF (dashed line), and the GVB approach (Projection Method, 
lower solid line) as a function of the applied magnetic field $B$. The two
other columns display the sS-UHF spin-up ($\uparrow$) and spin-down 
($\downarrow$) occupied orbitals (modulus square) of the singlet state for 
$B=0$ (field-free case, middle column) and $B=9$ T (right column). The top
half describes the case of a bare interdot barrier of $V_b=3.71$ meV, while
the bottom half describes the no barrier case $V_b=0$.  
For $B=9$ T complete dissociation has been practically reached.
The choice of the remaining parameters is: $m^*=0.067 m_e$, $\hbar \omega_0=5$
meV, $d=30$ nm, and $g^*=0$. Distances are in nm and the orbital densities 
in $10^{-4}$ nm$^{-2}$. Insets: The overlap integral (modulus square) of the 
two orbitals of the singlet state as a function of $B$.
}
\end{figure}
%***************** end figure 3 **************************
\subsubsection{Case of $\kappa=25$}

The influence of the magnetic field on the properties of the H$_2$-QDM in the 
case of a stronger $e-e$ repulsion ($\kappa=25$) is described in Fig.\ 3. 
Again we set $g^*=0$ for the same reasons as in the previous case; 
the meaning of $\Delta \varepsilon$ and of the displayed
orbitals is the same as in Fig.\ 2. Compared to the $\kappa=45$ case, the
increase of the $e-e$ repulsion is accompanied by an overall strengthening of
symmetry breaking and electron localization, since there is no range of 
parameters for which the sS-UHF and GVB solutions collapse into the 
symmetry-adapted MO (RHF) one. Indeed, even for $V_b=0$ the three levels of 
approximation provide different values for the singlet-triplet difference 
$\Delta \varepsilon$, with the GVB solution yielding in the field-free case
the highest stabilization for the singlet. 
For $V_b=3.71$ meV, the MO solution fails outright even in the field-free 
case for which it predicts a positive $\Delta \varepsilon$ (triplet ground 
state) in violation of the theorem stating \cite{matt} that the ground 
state of a 2$e$ system at $B=0$ is always a singlet.

The overall strengthening of electron localization is also reflected in
the behavior of the overlap integrals (see insets), since now they start
at $B=0$ with smaller values compared to the corresponding values of
the $\kappa=45$ case. In addition, the trend towards stronger electron
localization with smaller $\kappa$ can be further confirmed by an inspection
of the orbitals of the singlet state displayed in the middle ($B=0$) and right
($B=9$ T) columns of Fig.\ 3. Compared to the corresponding orbitals
in Fig.\ 2, this trend is obvious and we will not describe it in detail.
It will suffice to stress only that, whatever the degree of initial electron
localization at $B=0$, the applied magnetic field enhances even further this
localization, achieving again a practically complete dissociation
of the artificial molecule at $B=9$ T.

\newpage

\subsubsection{Case of $\kappa=12.9$ (GaAs)}

%***************** begin figure 4 **************************
\begin{figure}[t]
%\centering\includegraphics[width=12.5cm]{qds_lowdin_fig4j.eps}\\
%~~~~\\
\vspace{6.2cm}
\centering{\Large{\bf FIGURE 4}}
\vspace{6.5cm}
\caption{
Lateral H$_2$-QDM in the presence of a magnetic field and for $\kappa=12.9$
(GaAs): The left column displays the energy difference $\Delta \varepsilon$ 
between the singlet and triplet states according to the sS-UHF 
(long dashed line) and the GVB approach 
(Projection Method, solid line) as a function of the applied magnetic 
field $B$. The two other columns display the sS-UHF spin-up ($\uparrow$) and 
spin-down ($\downarrow$) occupied orbitals (modulus square) of the singlet 
state for $B=0$ (field-free case, middle column) and $B=9$ T (right column). 
The top half describes the case of a bare interdot barrier of $V_b=4.94$ meV, 
while the bottom half describes the no barrier case $V_b=0$.  
For $B=9$ T complete dissociation has been practically reached.
The choice of the remaining parameters is: $m^*=0.067 m_e$, $\hbar \omega_0=5$
meV, $d=30$ nm, and $g^*=-0.44$ (GaAs). Distances are in nm and the orbital 
densities in $10^{-4}$ nm$^{-2}$. Insets: The overlap integral 
(modulus square) of the two orbitals of the singlet state as a function of $B$.
}
\end{figure}
%***************** end figure 4 **************************
The dielectric constant for a GaAs heterointerface is $\kappa=12.9$, which
corresponds to a further increase in the $e-e$ repulsion compared to the
two previous cases of $\kappa=45$ and $\kappa=25$. This case is presented
in Fig.\ 4. Note that here the Zeeman contribution has been included
from the beginning by taken $g^*$ to be equal to its actual value in the GaAs
heterointerfaces, i.e., $g^*=-0.44$. As a result, the sS-UHF and GVB curves 
for $\Delta \varepsilon$ converge to a straight line representing the Zeeman 
linear dependence $\gamma B$ (with $\gamma \approx 0.026$ meV/T for $S_z=+1$, 
short dashed line) instead of vanishing for large $B$. 

The results presented in Fig.\ 4 confirm again the trend that electron
localization becomes stronger the stronger the interelectron repulsion.
As we found previously, the proper description of this strong electron 
localization requires consideration of symmetry breaking via the sS-UHF (long 
dashed curve in the $\Delta \varepsilon$ vs. $B$ plots) and the subsequent
construction of GVB wave functions via Projection Techniques (solid
curve in the  $\Delta \varepsilon$ vs. $B$ plots). The MO description is
outright wrong, since the corresponding $\Delta \varepsilon$ values are
positive in the whole interval $0 \leq B \leq 9$ T; in fact they are
so large that the whole MO curves lie outside the plotted ranges in Fig.\ 4.
For $B=9$ T, and in both the barrierless $(V_b=0)$ case and the case with an 
interdot barrier of $V_b=4.94$ meV, the artificial molecule has practically 
dissociated. Compared to the previous cases of $\kappa=45$ and $\kappa=25$, the
molecule starts with a stronger electron localization already in the
field-free case and with increasing $B$ it moves much faster towards complete 
dissociation.

\subsubsection{Overview of common trends in the singlet-triplet energy
difference}

According to the GVB calculations,
the common trend that can be seen in all cases (independent of the value
of $\kappa$) is that the singlet-triplet energy difference is initially 
negative (singlet ground state) in the field-free case. With increasing
magnetic field, this energy difference becomes larger (diminishes in
absolute value), crosses the value of zero at a certain $B_0$, and
remains positive (triplet ground state) for all $B \geq B_0$, converging
from above to the straight line representing the Zeeman contribution (or to 
zero if the Zeeman contribution is neglected).
This means that for sufficiently large values of $B$ the singlet-triplet 
energy difference is given simply by the Zeeman energy and that the molecule 
has practically dissociated. The trend towards dissociation is reached faster
the stronger the $e-e$ repulsion.

It is interesting to note again that for $\kappa=45$ (weaker $e-e$ repulsion)
the sS-UHF and GVB solutions collapse to the MO solution for values of the
magnetic field smaller than  $B \leq 2.9$ meV. However, for the stronger
$e-e$ repulsions ($\kappa=25$ and $\kappa=12.9$) the sS-UHF and GVB solutions 
remain energetically well below the MO solution. Since the separation
considered here ($d=30$ nm) is a rather moderate one (compared to the value 
$l_0=28.50$ nm at $B=0$ for the extent of the $1s$ lowest orbital of an
individual dot with $\hbar \omega_0=5$ meV), we conclude that there is 
a large range of materials parameters, interdot distances, and magnetic-field
values for which the QDM's are weakly coupled and cannot be described by the 
MO theory; a similar conclusion was reached in Ref.\ \cite{yl4} for the
field-free case.

\section{Conclusions}
\label{sec:5}

We have shown that, even in the presence of an applied magnetic field,
the sS-UHF method, in conjunction with the companion
step of the restoration of symmetries when such symmetries are broken, is
able to describe the full range of couplings in a QDM, from the
strong-coupling regime exhibiting delocalized molecular orbitals to the 
weak-coupling one associated with Heitler-London-type combinations
of atomic orbitals.

The breaking of space symmetry within the sS-UHF
method is necessary in order to properly describe the weak-coupling
and dissociation regimes of QDM's. The breaking of the space symmetry
produces optimized atomic-like orbitals localized on each individual dot. 
Further improvement is achieved with the help of Projection Techniques 
which restore the broken symmetries
and yield multideterminantal many-body wave functions. The method of the 
restoration of symmetry was explicitly illustrated for the case of the 
H$_2$-QDM in the presence of a magnetic field. It led to the introduction 
of a Generalized Valence Bond many-body wave function as the 
appropriate vehicle for the description of the weak-coupling and dissociation
regimes of artificial molecules. 

Additionally, we showed that the RHF, whose orbitals  preserve the space 
symmetries and are delocalized over the whole molecule, is naturally
associated with the molecular orbital theory. In a generalization of the 
field-free case of natural [36$-$38]  %%\cite{coul,murr,szab} 
and artificial \cite{yl4} molecules, it was found that the RHF fails to 
describe the weak-coupling and dissociation regimes of QDM's in the presence 
of an applied magnetic field as well. 

~~~~~~\\
~~~~~~\\
This research is supported by a grant from the U.S. Department of Energy
(Grant No. FG05-86ER45234).

\end{document}